\documentclass[a4paper,11pt]{article}
\pdfoutput=1
\usepackage{jcappub}
\usepackage{tikz,xcolor,hyperref}
\usepackage{amsmath, amssymb, amsthm, graphicx, epsfig, fancyhdr, slashed}
\usepackage[normalem]{ulem}
\usepackage{tikzsymbols}
\usepackage{natbib}
\usepackage{float}
\usepackage{orcidlink}
\usepackage[normalem]{ulem}
\usepackage{bm}
\usepackage{booktabs}
\usepackage{latexsym}
\usepackage{subfig}
\usepackage{amsfonts}
\usepackage{bigints}
\usepackage{braket}
\usepackage{multirow}
\usepackage{blindtext}
\usepackage{titlesec}
\useunder{\uline}{\ul}{}
 \usepackage{environ}
\usepackage{mathtools}
\usepackage{placeins}
\usepackage{relsize}

\newcommand{\gs}{g_\star}
\newcommand{\gss}{g_{\star s}}
\newcommand{\Trh}{T_\text{rh}}

\newcommand{\Tmax}{T_\text{max}}

\newcommand{\mdm}{m_\text{dm}}

\newcommand{\Tst}{T_{\star}}
\newcommand{\Hrad}{H_{\text{rad}}}

\begin{document}
%%%%%%%%%%%%%%%%%%
\title{Did LZ see modified gravity?}
%%%%%%%%%%%%%%%%%%%%%%%%%%%%%%%
\author{Basabendu Barman\,\orcidlink{0000-0003-0374-7655}}
\affiliation{Department of Physics, School of Engineering and Sciences, SRM University-AP, Amaravati 522240, India}
\emailAdd{basabendu.b@srmap.edu.in}
%%%%%%%%%%%%%%%%%
\abstract{We investigate whether the recently reported high-energy nuclear recoil event by the LUX-ZEPLIN (LZ) collaboration can be interpreted as a signal of freeze-in dark matter (DM) in a non-standard cosmological history. We consider DM production from the Standard Model thermal bath during an era of modified expansion preceding Big Bang nucleosynthesis. Requiring consistency with the observed DM abundance, cosmological and collider constraints, we identify viable parameter regions that can accommodate the LZ event. Our results demonstrate the potential of direct detection experiments to probe non-standard cosmological histories and, more broadly, alternative theories of gravity.}
%%%%%%%%%%%%%%
\maketitle
%%%%%%%%%%%
\section{Introduction}
\label{sec:intro}
%%%%%%%%%%%
The recent observation of an isolated high-energy nuclear recoil event by the LUX-ZEPLIN (LZ) experiment has generated considerable interest in possible new-physics interpretations. Using an exposure of $2.84~\mathrm{ton,yr}$, the LZ collaboration reported a nuclear recoil event with an energy of $E_R = 248 \pm 23~(\mathrm{stat}) \pm 23~(\mathrm{sys})~\mathrm{keV}$, which is in tension with the background-only hypothesis at a global significance of $2.6\sigma$~\cite{LZ:2026axp}. Interestingly, the relatively large recoil energy makes this event particularly well suited to scenarios in which dark matter (DM) undergoes inelastic scattering. In particular, the event can be accommodated by weakly interacting massive particle (WIMP) DM with masses above $\mathcal{O}(200~\mathrm{GeV})$, provided that the DM particle can scatter inelastically and transition to a heavier state through up-scattering on xenon nuclei. This observation has already motivated a number of DM interpretations, as well as alternative explanations. Several studies have explored Higgsino DM as a possible origin of the event~\cite{Fan:2026kxx,Wu:2026nhi,Freese:2026sga}, while other works have considered a variety of DM candidates and mechanisms, as well as non-DM possibilities~\cite{McCabe:2026crm,Unwin:2026rdp,Smirnov:2026aqk,Nomura:2026qyq,Lou:2026idn,Su:2026rwz,DiMauro:2026ldr,Yamashita:2026ump,Chattopadhyay:2026ryw,deLima:2026shq,Visinelli:2026kgt,Dent:2026bji,Gu:2026vto,Lee:2026xxh,Jeesun:2026vzo,Borah:2026zwf,Bandyopadhyay:2026gjw,Okada:2026eol,Okada:2026ypw,Okada:2026upm}. The minimal Higgsino interpretation, however, faces important phenomenological challenges. In particular, constraints from the non-observation of corresponding signals from DM captured in the Sun~\cite{Pospelov:2026ewn,Bose:2026ndd}, together with the absence of additional higher-energy recoil events in the LZ data~\cite{Rodd:2026tyn}, can place strong restrictions on the parameter space and may even disfavor the minimal scenario~\cite{Bose:2026ndd}. 

In canonical WIMP paradigm, DM interacts with the Standard Model (SM) plasma with interaction strengths of roughly electroweak size. These interactions keep DM in thermal equilibrium with the primordial plasma at early times, after which DM decouples through the freeze-out mechanism. However, increasingly stringent observational constraints (see, e.g., Refs.~\cite{Roszkowski:2017nbc, Arcadi:2017kky}) on the conventional WIMP parameter space have motivated the exploration of alternative DM production mechanisms. A well-motivated alternative is provided by feebly interacting massive particles (FIMPs)~\cite{McDonald:2001vt,Choi:2005vq,Kusenko:2006rh,Petraki:2007gq,Hall:2009bx,Bernal:2017kxu}. Unlike WIMPs, FIMPs are never in thermal equilibrium with the SM plasma. Instead, they can be produced through the decay or annihilation of particles in the visible sector. As the Universe cools below the relevant mass scale of the interaction, set by the larger of the DM and mediator masses, these production processes become Boltzmann suppressed. The comoving DM number density then approaches a constant value, giving rise to the freeze-in  mechanism~\cite{Hall:2009bx}. In the FIMP framework, the dark and visible sectors interact only very feebly, making the DM challenging to probe in different experimental searches. 

Motivated by these considerations, in this work we explore whether the recent LZ event can be explained within a FIMP DM framework. A central challenge, however, arises from the very feeble DM-SM interactions that are characteristic of FIMPs. While such small couplings are essential for keeping the DM population out of thermal equilibrium, they also suppress the DM-nucleus scattering rate, making it difficult to account for the observed recoil event within the standard cosmological history. This tension can be alleviated if DM is produced in a modified cosmological background in which the expansion rate of the Universe is larger than that during the standard radiation-dominated (RD) era. A faster expansion modifies the DM production dynamics and, for a fixed observed relic abundance, allows comparatively larger DM-SM couplings while the DM still remains out of thermal equilibrium [see, e.g.,~\cite{DEramo:2017ecx}]. Such an enhancement of the coupling can, in turn, increase the direct-detection scattering rate sufficiently to make a FIMP interpretation of the LZ event viable. The required departure from the standard cosmological history can arise naturally from modifications of Einstein's general relativity (GR), which alter the expansion history of the early Universe (see, e.g.,~\cite{Nojiri2007}). 

The paper is organized as follows. In Sec.~\ref{sec:framework}, we introduce the underlying particle physics framework. In Sec.~\ref{sec:DD}, we analyze the LZ event in the light of the present model. Sec.~\ref{sec:DM} discusses dark matter production in a modified cosmological background. Finally, we present our conclusions in Sec.~\ref{sec:concl}.
%%%%%%%%%%%
\section{The particle framework}
\label{sec:framework}
%%%%%%%%%%%
We consider a two-component inelastic DM scenario in which the dark sector consists of two Dirac fermions $\chi_H$ and $\chi_L$, split in mass by $\delta \equiv m_H - m_L \ll m_L$, which couple to a dark photon $X_\mu$ through an off-diagonal vector current. We shall discuss about a possible UV complete model for this set-up later. The dark photon in turn communicates with the SM via abelian kinetic mixing. The relevant interaction Lagrangian for this set-up is given by~\cite{Cui:2009xq},
\begin{equation}
\mathcal{L} \supset
- \frac{1}{4} X_{\mu\nu}X^{\mu\nu}
- \frac{\epsilon}{2} F_{\mu\nu} X^{\mu\nu}
+ \frac{1}{2} m_X^2 X_\mu X^\mu+g_D \, \bar{\chi}_H \gamma^\mu \chi_L \, X_\mu
+ \mathrm{h.c.}\,,
\label{eq:kinmix}
\end{equation}
where, $F^{\mu\nu}=\partial^{\mu}A^{\nu}-\partial^{\nu}A^{\mu}$ and $X^{\mu\nu}=\partial^{\mu}X^{\nu}-\partial^{\nu}X^{\mu}$ are the field-strength tensors of the SM and dark $U(1)$ gauge fields, $A^\mu$ and $X^\mu$, respectively, $g_D$ is the dark gauge coupling, and $\epsilon$ is the kinetic mixing parameter. Here we assume a Stueckleberg mass term\footnote{In abelian gauge theories, the Stueckelberg mechanism can be taken as the limit of the Higgs mechanism where the mass of the real scalar is sent to infinity and only the pseudoscalar is present~\cite{Stueckelberg:1938hvi,Ruegg:2003ps,Kors:2004dx,Kors:2005uz}.} for $X$. Diagonalizing Eq.~\eqref{eq:kinmix} one arrives at an interaction Lagrangian,
\begin{align}\label{eq:kinmix2}
& \mathcal{L}\supset-\frac{e\,\epsilon}{\sqrt{1-\epsilon^2}}\,J_\mu\,X^\mu+e\,J_\mu\,A^\mu\,,
\end{align}
where $J_\mu$ is the SM electromagnetic current. This term mediates both direct detection scattering as well as DM production in the early Universe. A simple renormalizable realization of the above effective theory can be obtained by embedding the dark sector in a spontaneously broken $U(1)_D$ gauge theory. We introduce two Dirac fermions $\psi_1$ and $\psi_2$ carrying opposite $U(1)D$ charges, together with a complex dark Higgs field $\Phi_D$. The dark Higgs generates both the dark photon mass and the mass mixing between $\psi_1$ and $\psi_2$. After spontaneous symmetry breaking and diagonalization of the fermion mass matrix, the resulting mass eigenstates can be identified with the two dark matter states $\chi_H$ and $\chi_L$. 
%%%%%%%%%%%
\section{The LZ event}
\label{sec:DD}
%%%%%%%%%%%
Since the interaction in Eq.~\eqref{eq:kinmix} only connects $\chi_H$ and $\chi_L$, elastic scattering is forbidden and the leading direct detection process is the inelastic up-scatter $\chi_H \, N \to \chi_L \, N$ off a target nucleus $N$, mediated by $t$-channel exchange of $X$. The resulting DM--proton cross section is given by~\cite{Essig:2011nj,Hambye:2018dpi},
\begin{equation}
\sigma_{\chi p}=\frac{16\pi \, \alpha_{\rm em} \, \alpha_D \, \epsilon^2 \, \mu_{\chi p}^2}
{\left(2 m_N E_R + m_X^2\right)^2}\,,
\label{eq:sigdd}
\end{equation}
where $\alpha_D \equiv g_D^2/4\pi$, $\alpha_{\rm em}$ is the fine-structure constant, $\mu_{\chi p} = m_H m_p /(m_H + m_p)$ is the reduced mass of the incoming state $\chi_H$ and the proton, $m_N$ is the nuclear mass, and $E_R$ is the nuclear recoil energy. The nuclear-level differential cross section, coherently enhanced by the nuclear charge $Z$, is given by~\cite{Lin:2019uvt}
\begin{equation}
\frac{d\sigma}{dE_R} =
\frac{m_N}{2\mu_{\chi p}^2 v^2} \,
\sigma_{\chi p}(E_R) \, Z^2 \, F^2(q r_A) ,
\label{eq:dsigdER}
\end{equation}
with $F(qr_A)$ the Helm nuclear form factor~\cite{Helm:1956zz} evaluated at momentum transfer $q = \sqrt{2 m_N E_R}$. The inelastic kinematics modify the minimum DM velocity required to produce a recoil $E_R$~\cite{Tucker-Smith:2001myb},
\begin{equation}
v_{\min}(E_R) = \frac{1}{\sqrt{2 m_N E_R}}
\left( \frac{m_N E_R}{\mu_{\chi N}} + \delta \right) ,
\label{eq:vmin}
\end{equation}
where $\mu_{\chi N} = m_H m_N/(m_H+m_N)$. Unlike the elastic case, $v_{\min}(E_R)$ is minimized (rather than monotonic) at $E_R^{*} = \delta \, \mu_{\chi N}/m_N$, which sets the recoil energy at which the differential rate is least kinematically suppressed. Requiring this characteristic recoil energy to coincide with the 248 keV event reported by LZ gives $\delta\simeq 280$ keV for a DM of mass 1 TeV. Thus, a mass splitting of order 280 keV naturally places the kinematic enhancement of the inelastic recoil spectrum near the observed event energy. Requiring only kinematic accessibility: $v_{\rm min}<v_{\rm esc}+v_\oplus$, instead gives the weaker upper bound $\delta\lesssim 360$ keV.

We calculate the nuclear-recoil event rate in a xenon-based detector within the standard halo model (SHM), incorporating the detector efficiency and finite energy resolution to obtain a prediction directly comparable with the observed events. The differential event rate per unit detector mass and recoil energy $E_R$ is
\begin{equation}
\frac{dR}{dE_R}=N_T\,n_{\rm DM}\,\int_{v > v_{\min}(E_R)}
\frac{d\sigma}{dE_R}v f_\oplus(\vec{v})\,d^3v\,,
\label{eq:dRdER_master}
\end{equation}
where $N_T$ denotes the number of target nuclei per unit detector mass, $n_{\rm DM} = \rho_{\rm DM}/m_{\rm DM}$ is the local dark matter number density, with $\rho_{\rm DM} = 0.3~\mathrm{GeV/cm^3}$, and $f_\oplus(\vec v)$ is the dark matter velocity distribution in the Earth frame. For $d\sigma/dE_R \propto 1/v^2$ [Eq.~\eqref{eq:dsigdER}], the velocity dependence is expressed through the mean inverse speed,
\begin{equation}
\eta(v_{\min}) \equiv \int_{v > v_{\min}} \frac{f_\oplus(\vec v)}{v}\, d^3v\,,
\label{eq:eta_def}
\end{equation}
giving
\begin{equation} \frac{dR}{dE_R} = N_T \, n_{\rm DM} \, \frac{m_N}{2\mu_{\chi p}^2} \, \sigma_{\chi p}(E_R) \, Z^2 \, F^2(qr_A) \; \eta(v_{\min}(E_R))\,. 
\label{eq:dRdER_final} 
\end{equation}
%%%%%%%%%%%%%%%%
\begin{figure}[htb!]
\centering
\includegraphics[scale=0.52]{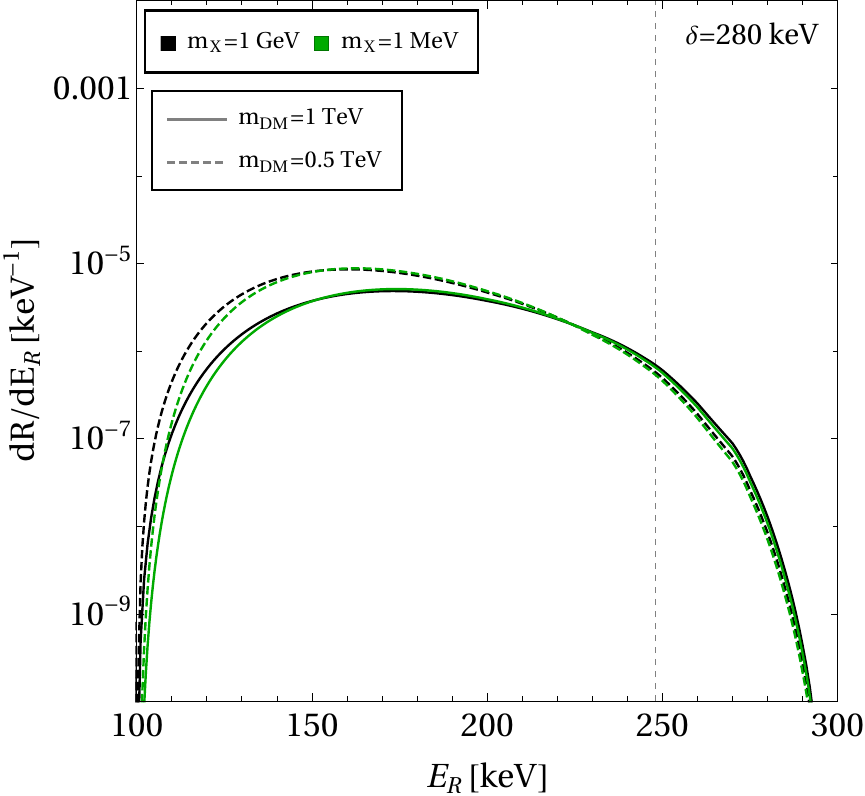}
\includegraphics[scale=0.52]{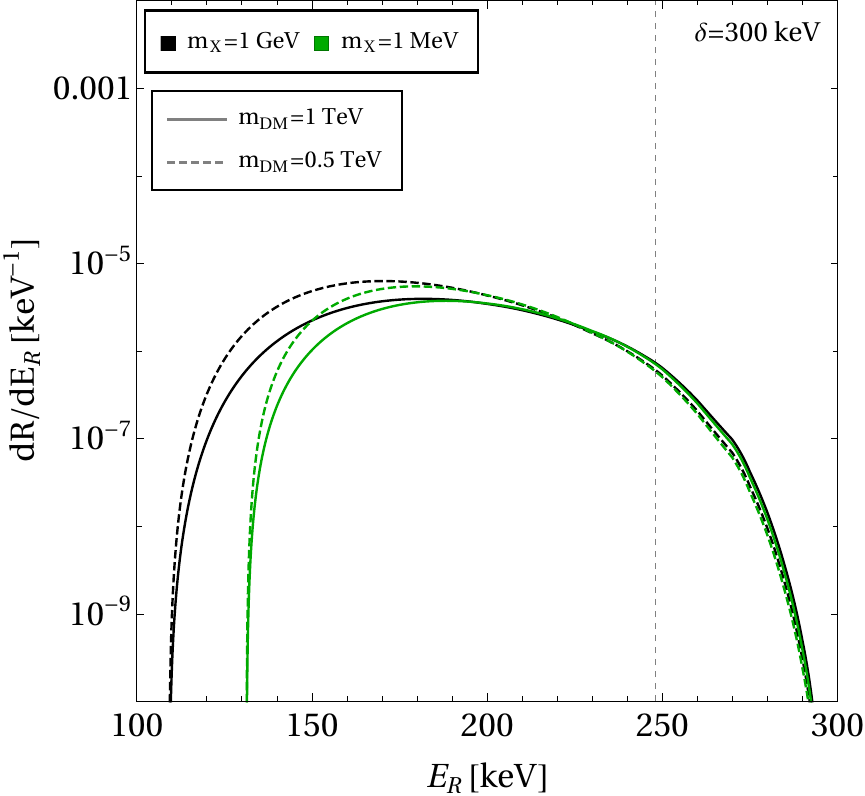}
\caption{Differential event rate as a function of the recoil energy $E_R$. The vertical dashed line corresponds to central recoil energy $E_R=248$ keV. All parameters are adjusted accordingly to obtain $N_{\rm sig}=1$ (see text for details).}
\label{fig:rate}
\end{figure}
%%%%%%%%%%%%%%%%%%%%%%%%%%%%%

We adopt a Maxwell-Boltzmann velocity distribution truncated at the galactic escape velocity and boosted by the Earth's velocity $v_\oplus$:
\begin{equation} f_\oplus(\vec v) = \frac{1}{N\,(\pi v_0^2)^{3/2}} \, e^{-|\vec v + \vec v_\oplus|^2/v_0^2} \, \Theta(v_{\rm esc} - |\vec v + \vec v_\oplus|)\,, \label{eq:fv} 
\end{equation}
where $v_0 = 220~\mathrm{km/s}$, $v_\oplus = 232~\mathrm{km/s}$, and $v_{\rm esc} = 544~\mathrm{km/s}$. The normalization factor is fixed by $\int f_\oplus\,d^3v = 1$,
\begin{equation} N =\mathrm{erf}\!\left(\frac{v_{\rm esc}}{v_0}\right) - \frac{2}{\sqrt{\pi}} \frac{v_{\rm esc}}{v_0} \, e^{-v_{\rm esc}^2/v_0^2} . \label{eq:Nesc} 
\end{equation}
After the angular integration, the mean inverse speed reduces to
\begin{equation} \eta(v_{\min}) = \frac{1}{N\, v_0^2\sqrt{\pi}\, v_\oplus} \int_{v_{\min}}^{v_{\rm esc}} v \left[ e^{-(v-v_\oplus)^2/v_0^2} - e^{-(v+v_\oplus)^2/v_0^2} \right] dv , \label{eq:eta_integral} \end{equation}
and vanishes for $v_{\min} \geq v_{\rm esc}$, thereby imposing the maximum kinematically accessible recoil energy. The nuclear response is described by the Helm form factor~\cite{Helm:1956zz},
\begin{equation} F^2(q) = \left[\frac{3\, j_1(q R_1)}{q R_1}\right]^2 e^{-q^2 s^2} , \qquad R_1 = \sqrt{R_A^2 - 5 s^2} , \quad R_A = 1.2\, A^{1/3}~{\rm fm}\,, \label{eq:helm} \end{equation}
where $s = 1~\mathrm{fm}$ is the skin thickness, $j_1$ is the spherical Bessel function of the first kind, and $q = \sqrt{2 m_N E_R}$ is the momentum transfer. To account for detector effects, the theoretical recoil spectrum is multiplied by the total signal efficiency $\varepsilon(E_R)$, which includes trigger, event-selection, and region of interest (ROI) efficiencies:
\begin{equation} \frac{dR}{dE_R}\to\varepsilon(E_R)\, \frac{dR}{dE_R}\,.
\label{eq:dRdER_obs} \end{equation}
For an exposure $\mathcal{E}$, defined as the fiducial detector mass times the live-time, the expected number of signal events in the analysis window $[E_-=5.4\,\text{keV},\, E_+=269.9\,\text{keV}]$ is,
\begin{equation} 
N_{\rm sig}=\mathcal{E} \int_{E_-}^{E_+} \frac{dR}{dE_R}\,dE_R\,. 
\label{eq:Nsig_benchmark} 
\end{equation}
Finally, to account for the finite energy resolution when comparing with a single reconstructed event at energy $E_c$ with resolution $\sigma$, we smear the predicted total count using a normalized Gaussian,
\begin{equation} \frac{dN_{\rm sig}}{dE_R} = \frac{N_{\rm sig}}{\sqrt{2\pi}\,\sigma} \, \exp\!\left[-\frac{(E_R - E_c)^2}{2\sigma^2}\right]\,, 
\label{eq:gaussian_smear} 
\end{equation}
which satisfies $\int\left(dN_{\rm sig}/dE_R\right)\,dE_R=N_{\rm sig}$ by construction. For LZ events, we consider $E_c=248$ keV, along with $\sigma=23\sqrt{2}$ keV. 

Throughout this work we identify viable model points by requiring the predicted signal count within the analysis window to satisfy $N_{\rm sig}=1$, motivated by the single candidate event observed by LZ in this window with negligible expected background, $b = 0.0106\pm0.0008$~\cite{LZ:2026axp}. This criterion is a reasonable and standard benchmarking device for scanning parameter space, but it does not constitute a rigorous statistical treatment of the data, for the following reasons. First, Eq.~\eqref{eq:Nsig_benchmark} fixes only the \emph{integrated} count and discards all information about \emph{where} the predicted events are expected to lie within the window. Two model points can both satisfy $N_{\rm sig}\simeq1$ while predicting drastically different recoil-energy spectra: one sharply peaked near the observed candidate's reconstructed energy $E_c = 248~\mathrm{keV}$, the other concentrated far from it, and Eq.~\eqref{eq:Nsig_benchmark} alone cannot distinguish between them. A model whose predicted spectral shape, once convolved with the detector's energy resolution $\sigma_E(E_R)$, places negligible probability density at $E_c$ is disfavored by the data even if it formally satisfies $N_{\rm sig}=1$. Second, requiring $N_{\rm sig}$ equal to unity treats the observed count as a point estimate rather than propagating the appropriate Poisson uncertainty. Relatedly, Eq.~\eqref{eq:Nsig_benchmark} neglects the (small but nonzero) expected background $b$, which should in principle be added to $N_{\rm sig}$ before comparison with the Poisson probability of observing one event. A statistically robust treatment instead requires an un-binned extended maximum-likelihood analysis of the form
\begin{equation}
\mathcal{L}(\vec\theta) = e^{-\left[N_{\rm sig}(\vec\theta)+b\right]}
\Big[N_{\rm sig}(\vec\theta)\,\mathcal{P}_{\rm sig}\!\left(E_c\,\big|\,\vec\theta\right) + b\,\mathcal{P}_{\rm bkg}(E_c)\Big] ,
\label{eq:likelihood}
\end{equation}
where $\vec\theta$ denotes the model parameters, $\mathcal{P}_{\rm sig}(E_R\,|\,\vec\theta)$ is the efficiency- and resolution-convolved signal probability density normalized to unit area over the analysis window, and $\mathcal{P}_{\rm bkg}(E_R)$ is the corresponding background spectral shape. Confidence regions on $\vec\theta$ would then follow from the profile of $-2\ln\mathcal{L}(\vec\theta)$, rather than from a single benchmark satisfying Eq.~\eqref{eq:Nsig_benchmark}. We adopt the simpler criterion of Eq.~\eqref{eq:Nsig_benchmark} in this work as it is sufficient to identify representative, order-of-magnitude-consistent parameter space and is standard practice in the phenomenological literature for rapidly surveying candidate explanations of a single anomalous event. We emphasize, however, that a definitive statistical statement regarding the compatibility of any specific model point with the LZ candidate event would require the full likelihood treatment of Eq.~\eqref{eq:likelihood}, incorporating the measured background spectral shape (see, e.g., Ref.~\cite{Bandyopadhyay:2026gjw}).

Fig.~\ref{fig:rate} shows the differential recoil rate spectra $dR/dE_R$ for several choices of the model parameters. The efficiency $\varepsilon(E_R)$ is included, but the spectra are not convolved with the Gaussian energy-resolution smearing function. The resulting recoil spectrum is worth commenting on. The distribution is broad, extending from approximately $100~\mathrm{keV}$ up to the upper end of the LZ ROI; it is not sharply peaked at $248~\mathrm{keV}$, though this value falls within the high-energy tail of the distribution. The feature near $270$--$290~\mathrm{keV}$ arises from the nature of the Helm form factor. Separately, for every benchmark choice of parameters we ensure that $N_{\rm sig} = 1$ within the ROI. As evident from Eq.~\eqref{eq:vmin}, increasing $\delta$ raises the minimum-velocity floor, reducing the fraction of the halo velocity distribution capable of inducing a scattering event. Consequently, for fixed coupling, the scattering rate is suppressed as $\delta$ increases. However, the curves shown are not computed at fixed coupling, rather all the remaining parameters for each choice of $\delta$ are fixed in a way so that $N_{\rm sig}=1$ (i.e., area under the curves is the same in all cases), as mentioned before. We see, for a DM of mass 1 TeV, we require $\epsilon\simeq\{2.2\times 10^{-5},\,6\times 10^{-8}\}$, corresponding to $m_X=\{1,\,10^{-3}\}$ GeV, where $\delta$ is fixed to 280 MeV, along with $\alpha_D=10^{-3}$. For a DM of mass 500 GeV, on the other hand, we fix $\epsilon\simeq\{1.3\times 10^{-6},\,10^{-7}\}$, corresponding to same choices of $m_X$. For a slightly larger $\delta=300$ keV, we obtain similar values. For the rest of the analysis we will consider two benchmark values of the DM mass: 500 GeV and 1 TeV. 
%%%%%%%%%%%
\section{Freeze-in in modified cosmology}
\label{sec:DM}
%%%%%%%%%%%
The DM in the present set-up is produced from the 2-to-2 scattering of the bath particles through: $f\bar{f}\to\chi_H \bar{\chi}_L$, where $f$ runs over kinematically accessible SM fermions. Since $\delta \ll m_H \simeq m_L \equiv \mdm$ is cosmologically negligible, we work in the degenerate-mass limit for the abundance calculation. The relevant cross section for DM production, summed over SM fermions with electric charge $Q_f$ and color multiplicity $N_{c,f}$, is given by
\begin{equation}
\sigma(s)=\sum_f N_c \,
\frac{16\pi \, \alpha_{\rm em} Q_f^2 \, \epsilon^2 \, \alpha_D}{3} \,
\frac{s\left(1 + \dfrac{2 m_{\rm DM}^2}{s}\right)
\sqrt{1 - \dfrac{4 m_{\rm DM}^2}{s}}}
{(s - m_X^2)^2 + m_X^2 \Gamma_X^2}\,,
\label{eq:sigmav_freezein}
\end{equation}
with $\alpha_D=g_D^2/(4\,\pi)$. The dark photon decays into the visible sector via Eq.~\eqref{eq:kinmix2} with decay rates~\cite{Fabbrichesi:2020wbt}
\begin{align}
&\Gamma_X=
\begin{dcases}
\displaystyle
\frac{N_c\,Q_f^2\,\alpha_{\rm em}\,\varepsilon^2}{3}\,m_{X}\,\sqrt{1-\frac{4m_\ell^2}{m_{X}^2}}\,\left(1+\frac{2m_\ell^2}{m_{X}^2}\right)\,, ~f\bar{f}\,,
\\[12pt]
\displaystyle
\frac{\mathbb{R}\,\alpha_{\rm em}\,\varepsilon^2}{3}\,m_{X}\,\sqrt{1-\frac{4m_\mu^2}{m_{X}^2}}\,\left(1+\frac{2m_\mu^2}{m_{X}^2}\right)\,,~\text{hadron}\,,
\end{dcases}
\end{align}
where $N_c=1(3)$ for leptons (quarks), $Q_f$ is the corresponding fermionic EM charge and $\mathbb{R}\equiv\sigma(e^+e^-\to\text{had})/\sigma(e^+e^-\to\mu^+\mu^-)$ is the EM spectral function ratio. 

In order to follow the evolution of the DM number density $n_{\rm DM}$, we express the Boltzmann equation in terms of the DM yield {\color{black}$Y_{\rm DM}\equiv n_{\rm DM}/\mathfrak{s}$} as,  
\begin{equation}\label{eq:beq}
x\,\mathcal{H}\,\mathfrak{s}\,\frac{dY_{\rm DM}}{dx} =\gamma(T)\,,
\end{equation}
where $x \equiv m_{\rm DM}/T$ is a dimensionless variable with $T$ the bath temperature. The reaction density $\gamma(T)$ is given by,
\begin{align}\label{eq:gamma22}
& \gamma(T) = \frac{T}{32\pi^4}\,g_a g_b 
\int_{\mathfrak{s}_{\rm min}}^\infty d\mathfrak{s}\,\frac{\big[(\mathfrak{s} - m_a^2 - m_b^2)^2 - 4m_a^2 m_b^2\big]}{\sqrt{\mathfrak{s}}}
\sigma(\mathfrak{s})_{a,b\to1,2}\,K_1\!\left(\frac{\sqrt{\mathfrak{s}}}{T}\right)\,,
\end{align}
where $g_{a,b}$ are the internal degrees of freedom of the initial states $a,b$, and the lower limit of the integration is $\mathfrak{s}_{\rm min} = \max\!\left[(m_a + m_b)^2,\,4m_\chi^2\right]$. The entropy density $\mathfrak{s}(T)$ and the Hubble rate $H_{\rm rad}(T)$ for a radiation-dominated Universe are given by,  
\begin{align}
&\mathfrak{s}(T)=\frac{2\pi^2}{45}\,\gss(T)\,T^3\,, & 
H_{\rm rad}(T)=\frac{\pi}{3}\,\sqrt{\frac{\gs(T)}{10}}\,\frac{T^2}{M_P}\, ,
\end{align}
where $\gss(T)$ and $\gs(T)$ denote the effective relativistic degrees of freedom associated with entropy and energy densities, respectively and $M_P$ is the reduced Planck mass. To account for the observed relic abundance, the present-day DM yield must satisfy  $Y_0\,m_{\rm DM} = \Omega h^2 \,\frac{1}{ \mathfrak{s}_0}\,\frac{\rho_c}{h^2}\simeq  4.3\times 10^{-10}\,{\rm GeV}$
where $Y_0 \equiv Y_{\rm DM}(T_0)$ is DM yield at present epoch, $\rho_c \simeq 1.05 \times 10^{-5} h^2\,{\rm GeV/cm}^3$ is the critical energy density, $ \mathfrak{s}_0\simeq 2.69 \times 10^3\,{\rm cm}^{-3}$ the present entropy density~\cite{Planck:2018vyg} and $\Omega h^2 \simeq 0.12$ the measured DM relic density~\cite{Planck:2018vyg}.

Modifications to GR can lead to cosmological histories in which the expansion rate of the Universe is larger than the Hubble expansion rate $H_\text{rad}$ predicted by the standard cosmological scenario. Such modified expansion histories can be conveniently parameterized as~\cite{Modak:1999nm, Schelke:2006eg, Catena:2009tm, Dent:2009bv, Leon:2013qh}
\begin{equation}\label{eq:hub-mod}
H(T)\equiv A(T)\,H_\text{rad}(T),,
\end{equation}
where $A(T)$ is the so-called amplification factor. The cosmological history before Big Bang Nucleosynthesis (BBN) is not directly constrained by observations. Therefore, for temperatures above $T_\text{BBN}$, the expansion rate can deviate significantly from the standard cosmological prediction, allowing $A(T)$ to differ from unity. On the other hand, the standard cosmological evolution must be recovered by the onset of BBN. Thus, $A(T)\neq1$ at early times, while $A(T)\to1$ as the Universe approaches the BBN epoch. A commonly used parametrization of the amplification factor is~\cite{Modak:1999nm, Schelke:2006eg, Catena:2009tm, Dent:2009bv, Leon:2013qh}\footnote{An alternative parametrization is given by~\cite{Catena:2009tm}
\begin{align}\label{A(T)} A(T)=
\begin{dcases}
1+\eta\left(\frac{T}{T_\star}\right)^n\,\tanh \frac{T-T_\text{re}}{T_\text{re}}\,, & T>T_\text{re}\,,
\\[12pt]
1\,, & T\leq T_\text{re}\,.
\end{dcases}
\end{align}
where $T_\star\geq T_\text{re}>T_\text{BBN}$. In the high-temperature limit, $T\gg T_\text{re}$, and for $n\geq0$, this parametrization reduces to that in Eq.~\eqref{eq:AT}.}
\begin{equation}\label{eq:AT}
A(T)=1+\eta\left(\frac{T}{T_\star}\right)^n,
\end{equation}
where $T_\star$ is a characteristic temperature scale, while $\eta$ and $n$ are dimensionless parameters that depend on the underlying cosmological model. Clearly, as $\Tst$ increases, the modified Hubble rate starts mimicking standard radiation domination, for $n>0$. The advantage of Eq.~\eqref{eq:AT} is that it also allows for negative values of $n$ while ensuring that the standard cosmological expansion is recovered for $T\leq T_\text{re}$. Different values of $\nu$ arise in various modified cosmological scenarios~\cite{Modak:1999nm, Schelke:2006eg, DAmico:2009tep, Catena:2009tm, Dent:2009bv, Leon:2013qh}. For example, $n=2$ corresponds to Randall--Sundrum type II brane cosmology~\cite{Randall:1999vf}, while $n=1$ is realized in kination scenarios~\cite{Salati:2002md, Pallis:2005hm, Guo:2009nt}. The case $n=0$ describes cosmologies with an approximately temperature-independent enhancement of the Hubble expansion rate, such as scenarios with a large number of additional relativistic degrees of freedom in the thermal plasma~\cite{Catena:2009tm}. In $f(x)$ cosmology with $f(x)=x+\alpha x^r$
where $x=R$ or ${\cal T}$, one finds $n=(2/r)-2$, with $R$ and ${\cal T}$ denoting the scalar curvature and scalar torsion, respectively~\cite{Capozziello:2008rq, Capozziello:2015ama, Cai:2015emx, Capozziello:2017bxm}.
%%%%%%%%%%%%%%%%
\begin{figure}[htb!]
\centering
\includegraphics[scale=0.6]{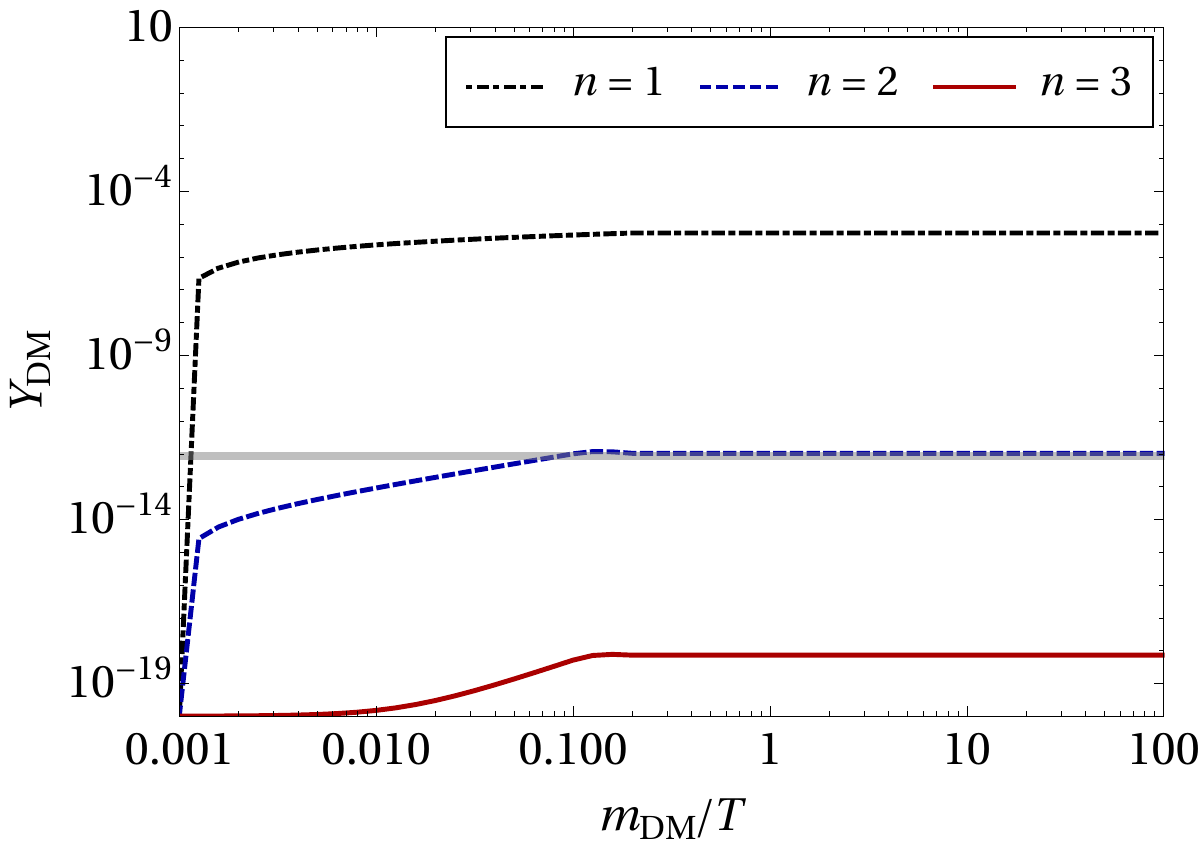}\\[12pt]
\includegraphics[scale=0.6]{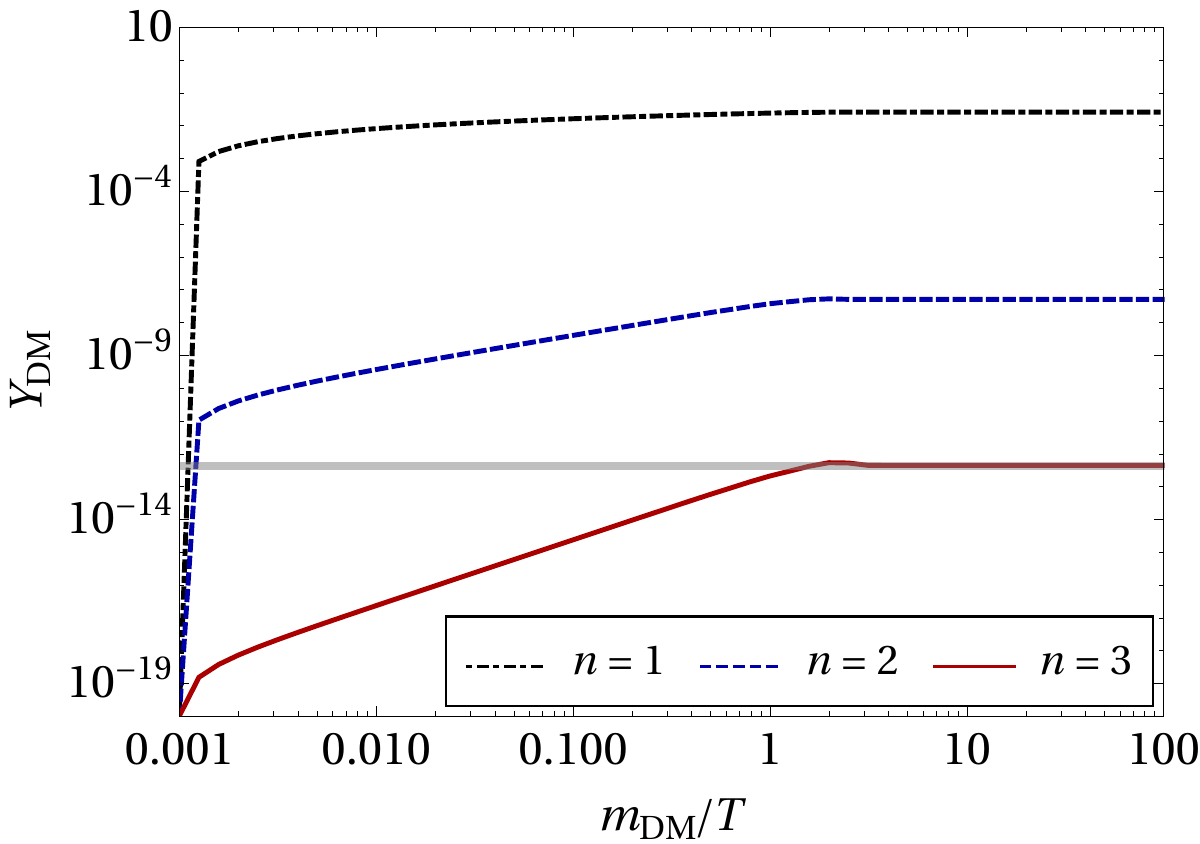}
\caption{DM yield as a function of the dimensionless quantity $\mdm/T$, for different choices of $n$, as shown by different curves. In the top panel we fix $\mdm=500\,\text{GeV},\,\Tst=5\,\text{MeV},\,\epsilon\simeq1.3\times 10^{-6},\,\eta\simeq 10^4,\,\alpha_D=10^{-3}$, while for the bottom panel $\mdm=1\,\text{TeV},\,\Tst\simeq11\,\text{MeV},\,\epsilon=2.2\times 10^{-5},\,\eta=10^3,\,\alpha_D=10^{-3}$, such that $N_{\rm sig}=1$ within ROI. In all cases we choose $m_X=1$ GeV.}
\label{fig:yld}
\end{figure}
%%%%%%%%%%%%%%%%%%%%%%%%%%%%%

Using the modified Hubble rate in Eq.~\eqref{eq:hub-mod}, the DM yield at a temperature $T$ can be obtained by solving the BEQ in Eq.~\eqref{eq:beq},
\begin{align}\label{eq:dm-yield}
& Y(T)=-\frac{135\,M_P\sqrt{5/2}}{\pi^3\,\gss\,\sqrt{\gs}}\,\int_{\Trh}^T\,dT\,\frac{\gamma(T)}{T^6}\,\left[1+\eta\,\left(\frac{T}{T_\star}\right)^n\right]^{-1}\,.
\end{align}
Here, $\Trh$ is the reheating temperature defined under the sudden inflaton-decay approximation, as the maximum temperature of the thermal bath\footnote{Away from the sudden decay approximation for reheating, the bath temperature may rise to a temperature $\Tmax\gg\Trh$~\cite{Giudice:2000ex,Kolb:2003ke}.}. In deriving Eq.~\eqref{eq:dm-yield}, a vanishing initial DM abundance has been assumed, as appropriate for freeze-in scenarios. To analytically understand the aspects of freeze-in, we work in the limit where the initial and final states have negligible masses with respect to the center of mass energy $\sqrt{s}$. This is a legitimate approximation for early times when $T\gg m$, typically above the EW symmetry breaking where all SM fields are massless. For the detailed analysis, we however solve the BEQ full numerically taking all masses and decay rates into account. In the massless limit, the total DM production cross-section simplifies to,
\begin{eqnarray}\label{eq:cs-approx}
\sigma(s)\simeq\frac{128\,\pi}{3}\,\alpha_{\rm em}\,\alpha_D\,\epsilon^2\,\frac{s}{(s-m_X^2)^2+m_X^2\,\Gamma_X^2}\,.
\end{eqnarray} 
In this limit, the reaction rate density in Eq.~\eqref{eq:gamma22} also simplifies to
\begin{align}\label{eq:gamma-approx}
&\gamma(T)\simeq g_a\,g_b\frac{T}{32\,\pi^4}\,\int_0^\infty ds\,s^{3/2}\,\sigma(s)\,K_1\left(\frac{\sqrt{s}}{T}\right)\,.   
\end{align}
Plugging Eq.~\eqref{eq:cs-approx} in Eq.~\eqref{eq:gamma-approx} one obtains,
\begin{align}
&\gamma(T)\propto\epsilon^2\,\alpha_D
\begin{dcases}
\frac{T^8}{m_X^4}\,, & T\ll m_X/2\,,
\\[12pt]
\frac{T\,m_X^4}{\Gamma_X}\,K_1\left(m_X/T\right)\,, & T\simeq m_X/2\,,
\\[12pt]
T^4\,, & T\gg m_X/2\,,
\end{dcases}
\end{align}
showing, for lighter mediator $(T\gg m_X/2)$, corresponding production rate is independent of the mediator mass; whereas for heavier mediator $(T\ll m_X/2)$, the quintessential UV nature of freeze-in prevails. With the approximations and assumptions, we obtain an approximate analytical expression for the DM yield via 2-to-2 scattering as,
\begin{align}
Y_0\propto M_P\,\epsilon^2\,\alpha_D
\begin{dcases}
\frac{\Trh^3}{m_X^4}\,, & T\ll m_X/2\,~~~~~(0<n<3)\,,
\\[12pt]
\frac{m_X^4}{\Gamma_X\,T_0^4}\,\left(\frac{\Tst}{T_0}\right)^n\,\left[1-\left(\frac{T_0}{\Trh}\right)^{n+4}\right]\,, & T\simeq m_X/2\,,
\\[12pt]
\frac{1}{T_0}\,\left(\frac{\Tst}{T_0}\right)^n\,\left[1-\left(\frac{T_0}{\Trh}\right)^{n+1}\right]\,, & T\gg m_X/2\,,
\end{dcases}
\end{align}
where we have considered $H(T)\approx\Hrad\,\eta\,(T/\Tst)^n$. For $n=3$, the final abundance turns out to be
\begin{align}
Y_0\propto\epsilon^2\,\alpha_D\,\frac{\Tst^3\,M_P}{m_X^4}\,\ln\left(\frac{\Trh}{T_0}\right)\,,    
\end{align}
for $T\ll m_X/2$, while for $T\gg m_X/2$, the final yield becomes independent of $\Trh\gg T_0$. 

Importantly, for freeze-in to be valid, the DM production rate must remain sub-Hubble until freeze-in completes. We check this by comparing the scattering rate $\gamma(T)/n_{\rm eq}^{\rm DM}$ with $H$, where the equilibrium DM number density is $n_{\rm eq}^{\rm DM} = \frac{2T}{\pi^2}\,m_\chi^2\,K_2\!\left(\frac{m_\chi}{T}\right)$, thereby demanding
\begin{align}
& \mathcal{R}\equiv\frac{\gamma}{n_{\rm eq}^{\rm DM}\,H}\Bigg|_{T=\mdm}<1\,.
\end{align}
In the present case, since $\mdm\gg m_X$ is required to explain the LZ event, hence we need to ensure that the DM remains out of equilibrium at $T\simeq\mdm$. Now, using Eq.~\eqref{eq:gamma-approx} and the modified Hubble rate we obtain an upper bound on the couplings as,
\begin{align}
& \alpha_D\,\epsilon^2\lesssim \frac{\eta}{M_P}\,\left(\frac{\mdm}{\Tst}\right)^n
\begin{dcases}
\frac{0.9\,m_X^4}{\mdm^3}\,, & T\ll m_X/2\,,
\\[12pt]
874\,\Gamma_X\,\frac{(\mdm/m_X)^4\,}{K_1\left(m_X/\mdm\right)}\,, & T\simeq m_X/2\,,
\\[12pt]
686\,\mdm\,, & T\gg m_X/2\,,
\end{dcases}
\end{align}
which clearly shows, for a given DM mass, a larger $\eta$ and a larger $n$ relaxes the bound on $\epsilon$, for a fixed $\Tst\ll\mdm$ and $\alpha_D$. We numerically obtain the maximum allowed $\epsilon$, for a given DM mass, mediator mass and $\alpha_D$, above which $\mathcal{R}>1$.

The evolution of the DM yield as a function of $\mdm/T$ is shown in Fig.~\ref{fig:yld} for different choices of the DM mass and couplings. In all cases, the parameters are chosen to reproduce the observed LZ event. For fixed DM mass and couplings, increasing $n$ leads to a smaller asymptotic yield and hence to an under abundant DM relic density. This can be understood from the enhanced Hubble expansion rate for larger $n$, which reduces the efficiency of DM production and dilutes the DM number density more rapidly. Therefore, obtaining the observed DM abundance requires larger couplings, which enhance DM production from the thermal bath. For the same choices of masses and couplings, the corresponding yield in the standard RD cosmology is much larger than the observed DM abundance (overproduction). We therefore do not show the RD results in these plots.
%%%%%%%%%%%%%%%%
\begin{figure}[htb!]
\centering
\includegraphics[scale=0.64]{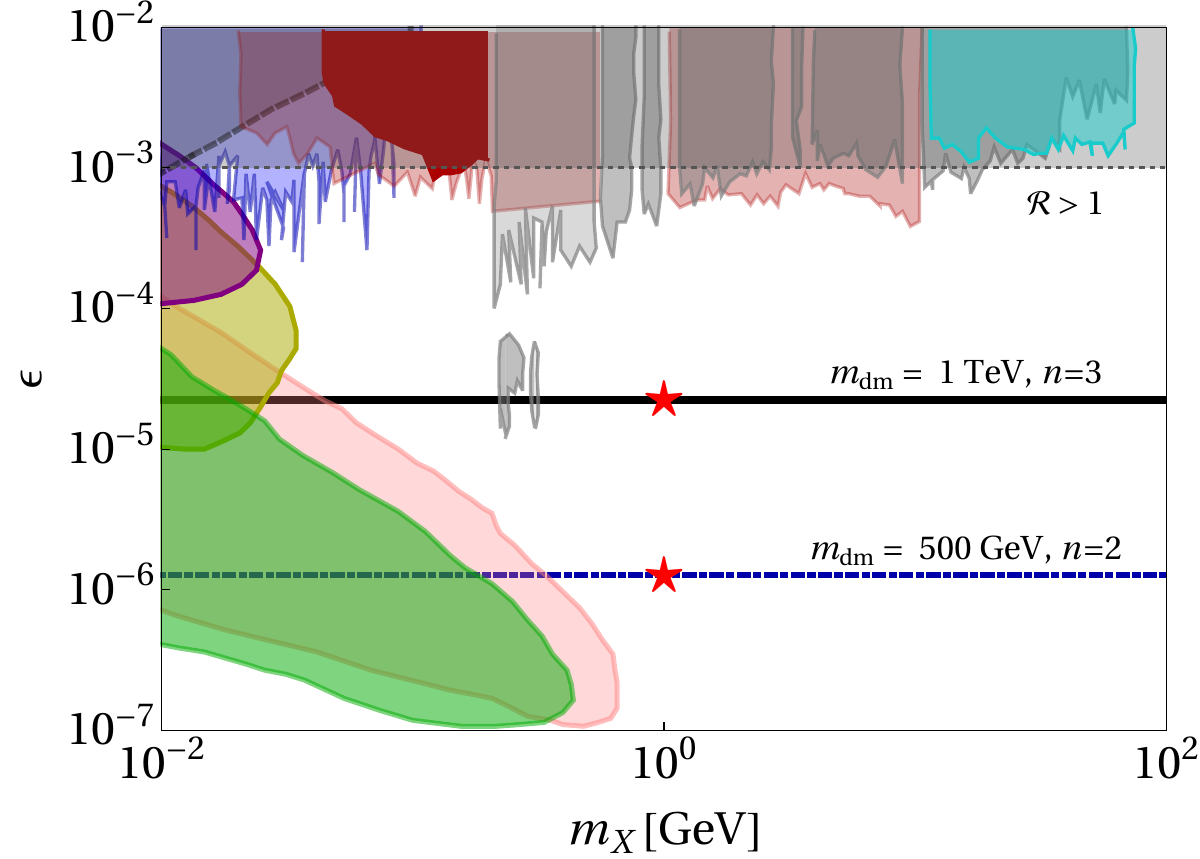}
\caption{Summary of the viable parameter space. The black solid and blue dashed horizontal lines provide the right DM abundance for two benchmark choices of the DM mass, as indicated. Existing limits from di-lepton searches  at low energy scattering, high energy collider and fixed target experiments: A1~\cite{Merkel:2014avp} (darker red),  LHCb~\cite{Aaij:2019bvg} (gray), 
CMS~\cite{CMS:2019kiy} (cyan),
BaBar~\cite{Lees:2014xha} (lighter red), 
NA48/2~\cite{Batley:2015lha} (blue), and beam–dump experiments:  E141~\cite{Riordan:1987aw} (yellow), $\nu$-Cal~\cite{Blumlein:2011mv,Blumlein:2013cua} (pink), CHARM~\cite{Gninenko:2012eq} (green), and projected limit from HL-LHC~\cite{Curtin:2014cca} (red, dashed) are shown. The red star marks the observed LZ event.}
\label{fig:sum}
\end{figure}
%%%%%%%%%%%%%%%%%%%%%%%%%%%%%

A summary of the parameter space is presented in Fig.~\ref{fig:sum}, in the bi-dimensional plane of kinetic mixing versus dark photon mass. The thick horizontal line corresponds to observed DM abundance for a DM of mass 1 TeV, shown by the black solid line and 500 GeV, indicated by the red broken line. Note that, the DM abundance is independent of the mediator mass, a nature that follows from the third line of Eq.~\eqref{eq:gamma-approx}. Also, for heavier DM masses, we find that larger values of $n$ are preferred for approximately the same $\Tst$. This can be understood from the fact that a faster expansion leads to a stronger dilution of the produced abundance. Therefore, a larger expansion rate can compensate for the larger relic abundance associated with heavier DM, allowing the observed DM abundance to be obtained. Due to $\mdm\gg m_X$, these contours are insensitive to choice of $\Trh$, a typical feature of IR freeze-in, as opposed to UV freeze-in. The $\star$ marks mediator mass that explains the observed LZ event, for a given DM mass. In order to achieve the right abundance for different DM masses, we fix all other parameters following Fig.~\ref{fig:yld}. It is worth mentioning that right DM abundance can be obtained by tuning $\Tst$ to a larger value, while at the same time increasing $\eta$, keeping all other parameters fixed. We also show existing limits from di-lepton searches  at low energy scattering, high energy collider and fixed target experiments: A1~\cite{Merkel:2014avp},  LHCb~\cite{Aaij:2019bvg}, CMS~\cite{CMS:2019kiy},
BaBar~\cite{Lees:2014xha}, 
NA48/2~\cite{Batley:2015lha}, as well as beam–dump experiments:  E141~\cite{Riordan:1987aw}, $\nu$-Cal~\cite{Blumlein:2011mv,Blumlein:2013cua}, CHARM~\cite{Gninenko:2012eq}, and projected limit from HL-LHC~\cite{Curtin:2014cca}. Additional limits arises from the electron anomalous magnetic moment $(g-2)_e$~\cite{Pospelov:2008zw} (gray diagonal dashed line). Above the gray dotted line, the DM equilibrates at $T=\mdm$. 

Overall, while part of the parameter space for both $\mdm=500$ GeV and $\mdm=1$ TeV are already constrained by LHCb, CHRAM and $\nu$-Cal, a significant portion of the parameter space still remains unconstrained by current experimental bounds. Therefore, improved sensitivity in dark photon searches could either probe or rule out the dark photon mediated explanation of the LZ events, as well as the need of a modified cosmological history before BBN. Similarly, improved sensitivity in future LZ searches could test the possibility of the same. Thus, whichever search achieves improved sensitivity first could provide a crucial test of these possible explanations.
%%%%%%%%%%%%%%%%%%%%%
\section{Conclusions}
\label{sec:concl}
%%%%%%%%%%%%%%%%%%%%
The search for a feebly interacting dark sector is albeit challenging but at the same time can provide a rich phenomenology in terms of both particle physics and cosmology. As we have discussed, a freeze-in explanation of the recent LZ event can not only indicate towards an alternative dark matter production mechanism, but can also be a test for modified gravity and pre-BBN cosmology. Interestingly, modified cosmologies arising from UV completions of GR can leave characteristic fingerprints on the expansion history of the early Universe. These fingerprints may not remain hidden for long: the resulting primordial gravitational wave (PGW) background can potentially be within the reach of current and upcoming GW detectors (see, e.g.,~\cite{Bernal:2020ywq}). More data from the LZ experiment, dark photon search experiments, together with a future detection of PGWs, would therefore offer a rather compelling litmus test of our framework, probing its particle physics and cosmological ingredients from two complementary directions.
%%%%%%%%%%%%%%%%%
\section*{Acknowledgments}
%%%%%%%%%%%%%%%%
The author would like to thank Arindam Das, Pankaj Borah and Nirmal Raj for fruitful discussions, and for providing valuable feedback on the manuscript.
%%%%%%%%%%%%%%%%
\section*{Note added:}
%%%%%%%%%%%%%%%%
While completing this manuscript, Ref.~\cite{Cabo-Almeida:2026uqw} appeared, where the authors also addressed the LZ event in the context of freeze-in DM production. However, their mechanism for achieving freeze-in at relatively strong couplings, thereby accounting for the LZ event, is  different from the mechanism considered here. 
%%%%%%%%%%%%%%%%%%%%
\bibliography{Bibliography}
\bibliographystyle{JHEP}
%%%%%%%%%%%%%%%%%%
\end{document}